\documentclass[twocolumn,showpacs,amsfonts,amsmath,amssymb,aps,prl]{revtex4}
\usepackage{bm}
\usepackage{amsfonts}
\usepackage{graphicx}
\begin{document}
\title{Particle beams guided by electromagnetic vortices:
New solutions of the Lorentz, Schr\"odinger, Klein-Gordon, and Dirac equations}
\author{Iwo Bialynicki-Birula}
\email{birula@cft.edu.pl} \affiliation{Center for Theoretical Physics, Polish
Academy of Sciences \\ Al. Lotnik\'ow 32/46, 02-668 Warsaw, Poland}
\begin{abstract}
It is shown that electromagnetic vortices can act as beam guides for charged
particles. The confinement in the transverse directions is due to the rotation
of the electric and magnetic fields around the vortex line. Large class of
exact solutions describing various types of relativistic beams formed by an
electromagnetic wave with a simple vortex line is found both in the classical
and in the quantum case. In the second case, the motion in the transverse
direction is fully quantized. Particle trajectories trapped by a vortex are
very similar to those in a helical undulator.
\end{abstract}


\maketitle

Electromagnetic waves with vortices have been extensively studied both
theoretically and experimentally. This field of research has became known as
singular optics \cite{so}. In this work I take these studies one step further
and analyze the motion of charged particles in the vicinity of a vortex line. I
shall consider the simplest possible solution of Maxwell equations with a
straight vortex line and show that this configuration of the electromagnetic
field acts as a perfect beam guide for charged particles. I study these
nonspreading beams in the classical case, when the relativistic particle
trajectory is determined by the Lorentz equations, and also in the quantum
case, when the wave function describing the beam obeys the Schr\"odinger,
Klein-Gordon, or the Dirac equation. In the classical and in the quantum case,
I exhibit analytic solutions that enable one to fully understand the intricate
dynamics of these beams.

The electric and magnetic field vectors of my model Maxwell field are
\begin{subequations}
\label{emfield}
\begin{eqnarray}
{\bm E}(x,y,z,t)= B_0\omega\left(f(x,y,z,t),g(x,y,z,t),0\right),\\
{\bm B}(x,y,z,t)= B_0k\left(-g(x,y,z,t),f(x,y,z,t),0\right),
\end{eqnarray}
\end{subequations}
where $B_0$ is the field amplitude measured in units of the magnetic field and
\begin{subequations}
\label{fg}
\begin{eqnarray}
f(x,y,z,t) &=& x\cos(\omega t_-)+y\sin(\omega t_-),\\
g(x,y,z,t) &=& x\sin(\omega t_-)-y\cos(\omega t_-),
\end{eqnarray}
\end{subequations}
where $t_-=t-z/c$. This configuration of the field is the simplest example of
the EM field with a vortex line \cite{bb0,atop}. The solution of the Maxwell
equations given by Eqs.~(\ref{emfield}) is not as artificial as it may look at
a first glance. It is a fairly good approximation (near the $z$-axis and not
far from the waist compared to the Raleigh range) to a realistic circularly
polarized paraxial Laguerre-Gauss beam with $n=0$ and $m=1$.

The Lorentz equations of motion
\begin{eqnarray}\label{lorentz0}
m\,{\ddot\xi}^{\mu}(\tau)= e\,f^{\mu\nu}(\xi(\tau)){\dot\xi}_{\nu}(\tau),
\end{eqnarray}
for a particle moving in the field (\ref{emfield}), expressed in terms of the
components $(\xi,\eta,\zeta,\theta)$ of the four-vector ${\xi}^{\mu}(\tau)$,
have the form (for the sake of brevity, I shall occasionally drop the
dependence on the proper time $\tau$)
\begin{subequations}
\label{lorentz}
\begin{eqnarray}
{\ddot\xi} &=& \omega_c\,\omega\, f(\xi,\eta,\zeta,\theta)
\left(\dot\theta-\dot\zeta/c\right),\label{l1}\\
{\ddot\eta} &=& \omega_c\,\omega\, g(\xi,\eta,\zeta,\theta)
\left(\dot\theta-\dot\zeta/c\right),\label{l2}\\
{\ddot\zeta} &=& \frac{\omega_c\,\omega}{c}
\left({\dot\xi}f(\xi,\eta,\zeta,\theta)+{\dot\eta}g(\xi,\eta,\zeta,\theta)\right),
\label{l3}\\
c\,{\ddot\theta}&=& \frac{\omega_c\,\omega}{c}
\left({\dot\xi}f(\xi,\eta,\zeta,\theta)+{\dot\eta}g(\xi,\eta,\zeta,\theta)\right),
\label{l4}
\end{eqnarray}
\end{subequations}
where the dots denote derivatives with respect to $\tau$ and $\omega_c=eB_0/m$
is the cyclotron frequency. These equations are nonlinear but they can be
explicitly solved owing to conservation laws.

By subtracting Eq.~(\ref{l3}) from Eq.~(\ref{l4}), one obtains
${\ddot\theta}-{\ddot\zeta}/c = 0$ and this leads to the first conserved
quantity
\begin{eqnarray}\label{const1}
 {\dot\theta}-{\dot\zeta}/c =
 \sqrt{1\!+\!({\dot\xi}^2+{\dot\eta}^2+{\dot\zeta}^2)/c^2}-\dot\zeta/c =
 \mathcal{E} = \text{const}_1.
\end{eqnarray}
Apart from the factor $mc^2$, this constant is the light-front energy --- the
conjugate variable to $t_+=t + z/c$
\begin{eqnarray}\label{e}
 \mathcal{E}=\frac{1-v_z/c}{\sqrt{1-{\bm v}^2/c^2}}
 = \frac{\sqrt{m^2c^4 +{\bm p}^2c^2} - p_zc}{mc^2}.
\end{eqnarray}
Without any loss of generality one may assume that $\theta(0) = 0 = \zeta(0)$
and then Eq.~(\ref{const1}) integrated with respect to $\tau$ yields
\begin{eqnarray}\label{effective}
 \theta - \zeta/c = \mathcal{E}\,\tau.
\end{eqnarray}
Thus, in this case, the proper time is proportional to the light-front
variable. The second constant of motion is obtained by combining
Eqs.~(\ref{l1}-\ref{l3}) and it reads
\begin{eqnarray}\label{const2}
{\dot\zeta} - \frac{1}{2c\mathcal{E}}({\dot\xi}^2+{\dot\eta}^2) =
\frac{1}{2}(\frac{1}{\mathcal{E}}-\mathcal{E}) = \text{const}_2.
\end{eqnarray}
Since the phase of the wave field changes in proper time with frequency
$\omega\mathcal{E}$, I shall incorporate $\mathcal{E}$ into $\omega$ and define
the effective frequency $\Omega=\mathcal{E}\omega$.

Owing to Eq.~(\ref{const1}), the transverse motion separates from the
longitudinal motion and the equations for $\xi$ and $\eta$ may be solved first.
This task is made easier by transforming Eqs.~(\ref{l1}-\ref{l2}) to the frame
rotating (in proper time) with the angular velocity $\Omega/2$ around the
$z$-axis which amounts to replacing $\xi$ and $\eta$ by the new variables
\begin{subequations}\label{new}
\begin{eqnarray}
\alpha(\tau) &=& \xi(\tau)\cos(\Omega\tau/2) + \eta(\tau)\sin(\Omega\tau/2),\\
\beta(\tau) &=& -\xi(\tau)\sin(\Omega\tau/2) + \eta(\tau)\cos(\Omega\tau/2).
\end{eqnarray}
\end{subequations}
The equations of motion for $\alpha$ and $\beta$ read
\begin{subequations}\label{eqm}
\begin{eqnarray}
{\ddot\alpha} &=& \Omega\,{\dot\beta} + (\Omega^2/4)\,\alpha +
\omega_c\Omega\,\alpha,\\
{\ddot\beta} &=& - \Omega\,{\dot\alpha} + (\Omega^2/4)\,\beta
-\omega_c\Omega\,\beta.
\end{eqnarray}
\end{subequations}
These equations result from the following Hamiltonian
\begin{eqnarray}\label{ham}
H &=& \frac{1}{2m}(p_\alpha^2 + p_\beta^2) + \frac{m\omega_c^2}{2}(\alpha^2 +
\beta^2)\nonumber\\
 &-& (\omega_c+\frac{\Omega}{2})\alpha\,p_\beta -
(\omega_c-\frac{\Omega}{2})\beta\,p_\alpha.
\end{eqnarray}
Despite the quadratic form of the Hamiltonian, it is not exactly a harmonic
oscillator --- the frequencies of the oscillations depend through $\mathcal E$
on the initial conditions. Still, this Hamiltonian can be expressed in terms of
the complex eigenmode amplitudes $a_\pm$ and $a_\pm^*$ (classical counterparts
of the annihilation and creation operators),
\begin{eqnarray}\label{ham1}
H = \Omega_+\,a_+^*a_+ - \Omega_-\,a_-^*a_-,
\end{eqnarray}
where $\Omega_\pm=\Omega\sqrt{1\pm\kappa}/2$ and $\kappa=4\omega_c/\Omega$ is a
dimensionless parameter that controls the particle behavior in the $xy$-plane.
The amplitudes $a_\pm$ have the form
\begin{subequations}\label{annih}
\begin{eqnarray}
 a_+ &=& \frac{1}{2}\sqrt{\frac{\kappa_+}{\gamma}}
 \left(\!p_\beta - \gamma\,\alpha - i\frac{p_\alpha +
 \gamma\,\beta}{\kappa_+}\right),\\
 a_- &=& \frac{1}{2}\sqrt{\frac{\kappa_-}{\gamma}}
 \left(\frac{p_\beta +
 \gamma\,\alpha}{\kappa_-} + i(p_\alpha - \gamma\,\beta)\right),
\end{eqnarray}
\end{subequations}
where $\gamma = \vert eB_0\vert$ and $\kappa_\pm=\sqrt{1\pm\kappa}$. The minus
sign in the diagonal form of the Hamiltonian (\ref{ham1}) indicates that the
beam dynamics in the transverse plane is governed by the same combination of
the attractive/repulsive oscillator forces and the Coriolis force as one
encounters for a particle in the Paul trap \cite{paul}, an electron Trojan wave
packet in an atom (cf. Eq.~(7) of Ref.~\cite{bke}), or Trojan asteroids in the
Sun-Jupiter system \cite{bke}.

The general solution for $\xi(\tau)$ and $\eta(\tau)$ is obtained by solving
Eqs.~(\ref{eqm}) in terms of eigenmodes and then undoing the rotation
(\ref{new}). The final expression for the motion of particles in the plane
perpendicular to the vortex line can be compactly written in the complex form
\begin{eqnarray}\label{sol}
\xi(\tau)+i\eta(\tau)=
e^{i\Omega\tau/2}\big((C+iD\kappa_+)\sin(\Omega_+\tau)\nonumber\\
-(B\kappa_-\!-iA)\sin(\Omega_-\tau)
-(D+iC\kappa_+)\cos(\Omega_+\tau)\nonumber\\
+(A\kappa_-\!-iB)\cos(\Omega_-\tau)\big),
\end{eqnarray}
where the constants $A,B,C,D$ depend on the initial values of the transverse
positions and velocities
\begin{subequations}\label{sol1}
\begin{eqnarray}
A=(\xi_0+\dot\eta_0/2\omega_c)/\kappa_-\,,\;\;
B=\dot\xi_0/2\omega_c\,,\\
C=(\eta_0+\dot\xi_0/2\omega_c)/\kappa_+\,,\;\;D={\dot\eta_0}/2\omega_c\,.
\end{eqnarray}
\end{subequations}
\begin{figure}
\centering
\includegraphics[width=0.35\textwidth]{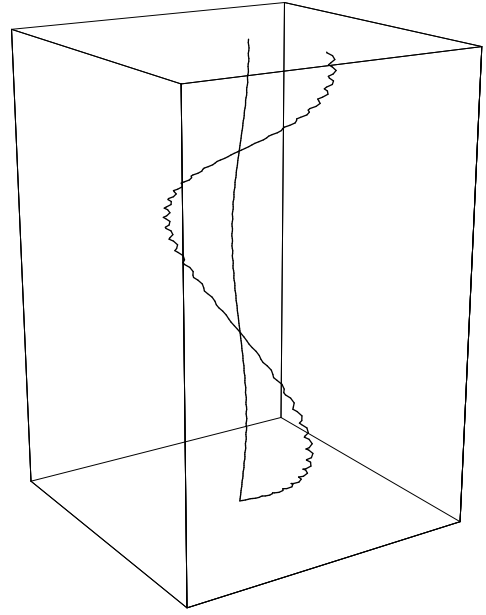}
\caption{Two trajectories of electrons injected into the wave field
(\ref{emfield}) with $B_0 = 10^{-3}\,\textrm{T}$ and $\omega=2\pi\cdot
10^9\,\textrm{s}^{-1}$. The initial longitudinal momentum $p_z$ of the electron
is in both cases $25\,\textrm{keV/c}$ but they have different transverse
momenta. The narrow trajectory has $p_x=5\,\textrm{keV/c}$ while the wide one
(wiggles on this trajectory are real) has $p_x=50\,\textrm{keV/c}$. The size of
the box measured in wavelengths $2\pi c/\omega$ is
$1\frac{1}{2}\times1\frac{1}{2}\times 2\frac{1}{2}$.}\label{fig}
\end{figure}
For $\vert\kappa\vert<1$ one obtains bounded oscillations around the vortex
line with four characteristic frequencies: $\Omega_+ \pm\Omega/2$ and
$\Omega_-\pm\Omega/2$ and for $\vert\kappa\vert>1$ one has runaway solutions
with exponential growth. The motion along the $z$-axis is obtained from
Eq.~(\ref{const2}) by a straightforward integration. The resulting formula for
${\zeta}(\tau)$ has two parts --- a part with oscillating terms and a linear
part in $\tau$ ,
\begin{eqnarray}\label{sol2}
\lefteqn{\zeta(\tau)=}\nonumber\\&&
\!\!\!\frac{\kappa^2\omega(D^2\!-\!C^2)\sin(2\Omega_+\tau)} {16c\kappa_+}+
\frac{\kappa^2\omega(B^2\!-\!A^2)\sin(2\Omega_-\tau)}
{16c\kappa_-}\nonumber\\&&
\!\!\!-\frac{\kappa^2\omega\,CD\!\left(1\!-\!\cos(2\Omega_+\tau)\right)}
{8c\kappa_+}+ \frac{\kappa^2\omega\,AB\!\left(1\!-\!\cos(2\Omega_-\tau)\right)}
{8c\kappa_-}\nonumber\\&&
 +\frac{c\tau}{2}\left(\frac{1}{\mathcal{E}}-\mathcal{E}+
\frac{\mathcal{E}\kappa^2\omega^2}{8c^2}(A^2+B^2+C^2+D^2)\right).
\end{eqnarray}

Depending on the sign of the linear term, the guiding center of the beam may
follow the electromagnetic wave or move in the opposite direction. By a special
choice of initial conditions one may even get rid of the linear term altogether
in which case the longitudinal motion will also be bounded, but it requires
fine tuning. This complex behavior is a purely relativistic effect. In the
nonrelativistic limit, the motion in the $z$-direction is free, not affected by
the wave at all, $\zeta(t)=v_zt$. In Fig.~\ref{fig}, I show two trajectories of
electrons for different initial conditions. These trajectories are very similar
to those in a helical undulator (an arrangement of permanent magnets used to
produce circularly polarized radiation). In the present case, the role of
permanent magnets is played by an electromagnetic wave with a vortex line and
the beam confinement is due to a totally different (Trojan) mechanism.

I shall start the analysis of the quantum-mechanical problem with the
Klein-Gordon (KG) equation. The EM field (\ref{emfield}) may be derived from
the vector potential
\begin{eqnarray}\label{empot}
{\bm A}(x,y,z,t)=B_0\left(g(x,y,z,t),f(x,y,z,t),0\right).
\end{eqnarray}
As seen from the analysis of the classical solutions, it is preferable to use
the coordinates $x,y$ and $t_\pm$. The KG equation in these coordinates reads
\begin{eqnarray}\label{kg}
\frac{4}{c^2}\partial_+\partial_-\psi = \left(\!\Delta_\perp -
\frac{e^2}{\hbar^2}{\bm A}^2 - 2i\frac{e}{\hbar}{\bm
A}\!\cdot\!{\bm\nabla}-\frac{m^2c^2}{\hbar^2}\right)\!\psi,
\end{eqnarray}
where $\partial_\pm = \partial/\partial t_\pm$. Since the variable $t_+$ does
not appear in this equation, one may seek its solutions in the form
\begin{eqnarray}\label{form}
\psi(x,y,t_-,t_+) = e^{-ic^2(Mt_+ + m^2t_-/M)/2\hbar}{\tilde\psi}(x,y,t_-).
\end{eqnarray}
An additional phase factor, dependent on $t_-$, has been introduced  to remove
the mass term. The essential dependence on $t_-$ is still contained in the wave
function ${\tilde\psi}$. The function ${\tilde\psi}(x,y,t_-)$ obeys the
following equation
\begin{eqnarray}\label{kg1}
i\hbar\partial_-{\tilde\psi} &=& \left(-\frac{\hbar^2}{2M}\Delta_\perp+
\frac{M\Omega_c^2}{2}(x^2+y^2)\right){\tilde\psi}\nonumber\\
&+&i\hbar\Omega_c\left((x\partial_y+y\partial_x)\cos(\omega
t_-)\right){\tilde\psi}\nonumber\\
&-&i\hbar\Omega_c\left((x\partial_x-y\partial_y)\sin(\omega t_-))\right)
{\tilde\psi},
\end{eqnarray}
where $\Delta_\perp$ is the transverse part of the Laplacian and
$\Omega_c=eB_0/M$. This equation is {\em exactly the same} as a nonrelativistic
Schr\"odinger equation except that the role of the mass $m$ is played by the
separation constant $M$ and the time parameter is replaced by the light-front
variable $t_-$. Therefore, everything that one can say about the solutions of
the Eq.~(\ref{kg1}) applies to the solutions of the Schr\"odinger equation.
Upon transforming the equation (\ref{kg1}) to a comoving frame by the
substitution
\begin{eqnarray}\label{subst}
{\tilde\psi}=\exp(-\frac{\omega t_-}{2}(x\partial_y-y\partial_x))\phi,
\end{eqnarray}
one finally obtains
\begin{eqnarray}\label{kg2}
\lefteqn{i\hbar\partial_-\phi =
\left(-\frac{\hbar^2}{2M}\Delta_\perp+\frac{M\Omega_c^2}{2}(x^2+y^2)\right)
\phi}\nonumber\\&&
\!\!\!+i\hbar\left((\Omega_c+\omega/2)x\partial_y\;+
(\Omega_c-\omega/2)y\partial_x\right)\phi\,.
\end{eqnarray}
By rearranging the terms, one may establish that the particle in this frame
moves effectively under the influence of the constant magnetic field ${\bm
B}=(0,0,m\omega/e)$ and an additional repulsive quadratic potential $V =
-(M\omega^2/8)(\kappa_+^2x^2+\kappa_-^2y^2)$.

All stationary solutions of Eq.~(\ref{kg2}) are most easily classified with the
use of the creation and annihilation operators. These operators diagonalize the
Hamiltonian
\begin{eqnarray}\label{ham2}
{\hat H} = ({\hat p}_x^2+{\hat p}_y^2)/2M +M\Omega_c^2({\hat
x}^2+{\hat y}^2)/2\nonumber\\
-(\Omega_c+\omega/2){\hat x}{\hat p}_y - (\Omega_c-\omega/2){\hat y}{\hat p}_x
\end{eqnarray}
and are obtained from the classical amplitudes (\ref{annih}) by the
replacements
\begin{subequations}
\begin{eqnarray}\label{repl}
a_\pm \to \sqrt{\hbar}{\hat a_\pm}\,,\;\;\;a_\pm^* \to \sqrt{\hbar}{\hat
a_\pm}^\dagger\,,\\
(\alpha,\beta,p_\alpha,p_\beta) \to ({\hat x},{\hat y},{\hat p}_x,{\hat
p}_y)\,.
\end{eqnarray}
This leads to the following form of the Hamiltonian
\end{subequations}
\begin{eqnarray}\label{ham3}
{\hat H} = \hbar\omega\left(\kappa_+({\hat a}_+^*{\hat a}_+ + 1/2) -
\kappa_+({\hat a}_-^*{\hat a}_- + 1/2)\right).
\end{eqnarray}
Thus, in contrast to the Volkov solution in the plane wave EM field
\cite{volkov}, the motion in the transverse direction is fully quantized. In
contrast to the motion in a constant magnetic field, the particle is localized
near the $z$-axis. Different normalization of the classical and quantum
Hamiltonian is due to the fact that the first one generates the evolution in
proper time, while the second one generates the evolution in the $t_-$
variable. These two parameters differ by the scaling factor $\mathcal E$. The
quantum theory becomes consistent with the classical one when $M/m$ is
identified with $\mathcal E$. It means that $M/m=\Omega/\omega$ and, as a
result, the value of $\kappa$ encountered in quantum theory becomes equal to
the classical one ($4\omega_c/\Omega=4\Omega_c/\omega$), as it should be.

Having diagonalized the Hamiltonian, one may generate the whole Fock space of
stationary solutions. They are obtained by acting on the fundamental state
$\phi_0$ with the creation operators. The fundamental state is the one
annihilated by both operators ${\hat a_\pm}$. Solving two simple differential
equations ${\hat a_\pm}\phi_0=0$, one obtains
\begin{eqnarray}\label{fund}
\phi_0(x,y) = N\exp(-x^2/d_+^2 - y^2/d_-^2 -ixy/d^2),
\end{eqnarray}
where the parameters $d_\pm$ and $d$ are given by
\begin{eqnarray}\label{param}
d_\pm^2 = \frac{\hbar}{\gamma}\frac{1+\kappa_+\kappa_-}{\kappa_\pm},\;\; d^2 =
\frac{\hbar}{\gamma}\frac{1+\kappa_+\kappa_-}{1-\kappa_+\kappa_-}\,.
\end{eqnarray}
The wave functions of the Fock states are polynomials in $x$ and $y$ multiplied
by the Gaussian ($\ref{fund}$). In the laboratory frame these solutions are not
stationary since the beams do not exhibit rotational symmetry around the
$z$-axis ($d_+\neq d_-$). In particular, the fundamental solution takes on a
form of a rotating helix.

There is also a plethora of nonstationary solutions of Eq.~(\ref{kg2}). First,
there are those that correspond directly to classical trajectories
--- the analogs of coherent states.
The fundamental solution (\ref{fund}) corresponds to a trajectory which just
sits on the vortex line, but one may easily obtain solutions of the KG equation
representing {\em all other} classical trajectories. According to a general
scheme \cite{bb1} valid for all quadratic Hamiltonians, displacing any solution
of the KG equation by the solutions of the classical equations of motion leads
to new solutions. Applying such displacements to the solution (\ref{fund}), one
obtains
\begin{eqnarray}\label{coh}
\phi(x,y,t_-) &=& N(\tau)e^{i(x p_\alpha(\tau)+y
 p_\beta(\tau))/\hbar}\nonumber\\
 &\times&\phi_0(x-\alpha(\tau), y-\beta(\tau)),
\end{eqnarray}
where $\tau= t_-/\mathcal E$ and the center-of-mass trajectories are obtained
by solving the Hamilton's equations of motion that follow from (\ref{ham}). The
time-dependent phase of the normalization constant is equal to the classical
action \cite{bb1}. To obtain the solution of the original equation, one must
transform the wave function from the comoving frame back to the laboratory
frame applying the inverse transformation to (\ref{subst}). Only then one
obtains the quantum-mechanical counterparts of the classical trajectories.

Solutions based on a rigid Gaussian
--- the analogs of coherent states --- do not exhaust all possibilities.
Since the center of mass motion of the Gaussian wave-packet decouples from its
internal motion, one may easily generate solutions based on oscillating
Gaussians
--- the analogs of squeezed states.
The Gaussian parameters $d_\pm$ and $d$ for such states are functions of $t_-$.
These states do not have direct classical counterparts and their complete
analysis will be given elsewhere.

The solution of the Dirac equation proceeds along similar lines. I begin with
rewriting the Dirac equation in the electromagnetic field (\ref{emfield})
\begin{eqnarray}\label{dirac0}
i\hbar\partial_t\Psi = \left(c{\bm\alpha}\!\cdot\!(-i\hbar{\bm \nabla} - e{\bm
A}) + \beta mc^2\right)\Psi,
\end{eqnarray}
as a set of two coupled equations for the two-component wave functions
\begin{subequations}\label{dirac1}
\begin{eqnarray}
2i\hbar\partial_+\Psi_+ =
c\left(mc\sigma_z-{\bm\sigma}_\perp\!\cdot\!(i\hbar\nabla+e{\bm
A})\right)\Psi_-,\label{dirac1a}\\
2i\hbar\partial_-\Psi_- =
c\left(mc\sigma_z-{\bm\sigma}_\perp\!\cdot\!(i\hbar\nabla+e{\bm
A})\right)\Psi_+,\label{dirac1b}
\end{eqnarray}
\end{subequations}
obtained with the use of the projections $P_\pm=(1\pm \alpha_z)/2$
\begin{eqnarray}\label{proj}
\Psi = (P_+ + P_-)\Psi = \Psi_+ + \Psi_-.
\end{eqnarray}
The dependence on the variable $t_+$ can again be separated by the same
substitution (\ref{subst}), leading to
\begin{eqnarray}
Mc{\tilde\Psi}_+ =
\left(mc\sigma_z-{\bm\sigma}_\perp\!\!\cdot\!(i\hbar\nabla+e{\bm
A})\right)\!{\tilde\Psi}_-,\label{diraca}
\end{eqnarray}
\begin{eqnarray}
\!\!\left(\!2i\hbar\partial_- + \!\frac{m^2c^2}{M}\right){\tilde\Psi}_-
=c\left(mc\sigma_z-{\bm\sigma}_\perp\!\!\cdot\!(i\hbar\nabla+e{\bm
A})\!\right)\!{\tilde\Psi}_+,\label{diracb}\nonumber\\
\end{eqnarray}
The first equation enables one to express ${\tilde\Psi}_+$ in terms of
${\tilde\Psi}_-$ and leads to Eq.~(\ref{kg1}) for ${\tilde\Psi}_-$. Again, as
in the case of the KG equation, the dependence of the potential on $t_-$ may be
eliminated by the substitution (\ref{form}) and the equation for $\Phi_-$ can
be reduced to {\em the same} Eq.~(\ref{kg2}) as for a spinless particle. Still,
the spin does play a role in the Dirac particle dynamics. Since the
transformation to the comoving frame should also involve the spin part, the
proper transformation rule, instead of (\ref{subst}), is now
\begin{eqnarray}\label{subst1}
{\tilde\Psi}_{\pm}=\exp(-\frac{\omega
t_-}{2}(x\partial_y-y\partial_x+i\sigma_z/2))\Phi_{\pm}.
\end{eqnarray}
Finally, the wave equation for ${\Phi}_-$ in the comoving frame
\begin{eqnarray}\label{dirac3}
i\hbar\partial_-\Phi_- \!=
\left(\!-\frac{\hbar^2}{2M}\Delta_\perp+\frac{M\Omega_c^2}{2}(x^2+y^2)
-\frac{\hbar\omega}{4}\sigma_z\!\right)\!\!\Phi_-\nonumber\\
+ i\hbar\left((\Omega_c+\frac{\omega}{2})x\partial_y\;+
(\Omega_c-\frac{\omega}{2})y\partial_x\right)\!\!\Phi_-\;\;\;\;
\end{eqnarray}
differs from Eq.~(\ref{kg2}) only by a simple spin term. Everything that has
been said before about stationary solutions of the KG equation applies with
almost trivial changes to the Dirac equation.

There are two properties of the solutions of the wave equations described here
that might lead to new effects: the quantization of the transverse motion and
the breaking of the rotational symmetry. This may help to observe the
rotational frequency shift predicted some time ago \cite{bb2} that depends
crucially on these features.

I would like to thank Zofia Bialynicka-Birula and Tomasz Rado{\.z}ycki for
critical comments and helpful advice. This work was supported by a grant from
the State Committee for Scientific Research in the years 2004-05.

\end{document}